\documentclass[lettersize,journal]{IEEEtran}
\pdfoutput=1
\makeatletter
\long\def\@makecaption#1#2{\ifx\@captype\@IEEEtablestring%
\footnotesize\begin{center}{\normalfont\footnotesize #1}\\
{\normalfont\footnotesize\scshape #2}\end{center}%
\@IEEEtablecaptionsepspace
\else
\@IEEEfigurecaptionsepspace
\setbox\@tempboxa\hbox{\normalfont\footnotesize {#1.}~~ #2}%
\ifdim \wd\@tempboxa >\hsize%
\setbox\@tempboxa\hbox{\normalfont\footnotesize {#1.}~~ }%
\parbox[t]{\hsize}{\normalfont\footnotesize \noindent\unhbox\@tempboxa#2}%
\else
\hbox to\hsize{\normalfont\footnotesize\hfil\box\@tempboxa\hfil}\fi\fi}
\makeatother

\IEEEoverridecommandlockouts

\usepackage{cite}
\usepackage{amsmath,amssymb,amsfonts}
\usepackage{graphicx}
\usepackage{textcomp}
\usepackage{multirow}
\usepackage{subfigure} 
\usepackage{stfloats}
\def\BibTeX{{\rm B\kern-.05em{\sc i\kern-.025em b}\kern-.08em
    T\kern-.1667em\lower.7ex\hbox{E}\kern-.125emX}}
\usepackage[T1]{fontenc} 

\usepackage{balance}

\usepackage{listings}
\usepackage{xcolor}

\definecolor{codegreen}{rgb}{0,0.6,0}
\definecolor{codegray}{rgb}{0.5,0.5,0.5}
\definecolor{codepurple}{rgb}{0.58,0,0.82}
\definecolor{backcolour}{rgb}{0.95,0.95,0.92}
\definecolor{GhostWhite}{rgb}{248 248 255}

\lstdefinestyle{mystyle}{
    backgroundcolor=\color{GhostWhite},   
    commentstyle=\color{codegreen},
    keywordstyle=\color{magenta},
    numberstyle=\tiny\color{codegray},
    stringstyle=\color{codepurple},
    basicstyle=\ttfamily\footnotesize,
    breakatwhitespace=false,         
    breaklines=true,                 
    captionpos=b,                    
    keepspaces=true,                 
    numbersep=5pt,                  
    showspaces=false,  
    frame = single,
    showstringspaces=false,
    showtabs=false,   
    rulesepcolor= \color{gray},
    tabsize=2
}

\usepackage{hyperref}
\renewcommand{\ref}[1]{\autoref{#1}}

\usepackage{algpseudocode}

\usepackage{algorithmicx,algorithm}

\usepackage{bbding}
\usepackage{color}

\newcommand{\citing}[1]{\textsuperscript{\cite{#1}}}

\begin{document}

\title{SNPSFuzzer: A Fast Greybox Fuzzer for Stateful Network Protocols using Snapshots}

\author{Junqiang Li, Senyi Li, Gang Sun, \textit{Member}, \textit{IEEE}, Ting Chen, \textit{Member}, \textit{IEEE} and Hongfang Yu, \textit{Member}, \textit{IEEE}
\thanks{This work was supported in part by the Chinese National Key Laboratory of Science and Technology on Information System Security (6142111200303), and in part by Key Research and Development Projects of Sichuan Province. (Corresponding author: Hongfang Yu.) 

Junqiang Li and Senyi Li are with School of Information and Communication Engineering, University of Electronic Science and Technology of China, Chengdu 611731, China (e-mail: lijunqiang@std.uestc.edu.cn; lisy@std.uestc.edu.cn).

Gang Sun is with Key Lab of Optical Fiber Sensing and Communications (Ministry of Education), University of Electronic Science and Technology of China, Chengdu 611731, China; and he is also with Agile and Intelligent Computing Key Laboratory of Sichuan Province, Chengdu 611731, China (e-mail: gangsun@uestc.edu.cn).

Ting Chen is with Center for Cybersecurity, University of Electronic Science and Technology of China, Chengdu 611731, China (e-mail: brokendragon@uestc.edu.cn).

Hongfang Yu is with Key Lab of Optical Fiber Sensing and Communications (Ministry of Education), University of Electronic Science and Technology of China, Chengdu 611731, China; and is also with Peng Cheng Laboratory, Shenzhen 518055, China (e-mail: yuhf@uestc.edu.cn).
}}

\markboth{}%
{}

\maketitle




\begin{abstract}
Greybox fuzzing has been widely used in stateless programs and has achieved great success. However, most state-of-the-art greybox fuzzers generally have the problems of slow speed and shallow state depth coverage in the process of fuzzing stateful network protocol programs which are able to remember and store details of the interactions.
 The existing greybox fuzzers for network protocol programs send a series of well-defined prefix sequences of input messages first and then send mutated messages to test the target state of a stateful network protocol. The process mentioned above causes a high time cost. In this paper, we propose SNPSFuzzer, a fast greybox fuzzer for stateful network protocol using snapshots. SNPSFuzzer dumps the context information when the network protocol program is under a specific state and restores it when the state needs to be fuzzed. Furthermore, we design a message chain analysis algorithm to explore more and deeper network protocol states. Our evaluation shows that, compared with the state-of-the-art network protocol greybox fuzzer AFLNET, SNPSFuzzer increases the speed of network protocol fuzzing by 112.0\%-168.9\% and improves path coverage by 21.4\%-27.5\% within 24 hours. Moreover, SNPSFuzzer exposes a previously unreported vulnerability in program Tinydtls. 
\end{abstract}


\section{Introduction}
\IEEEPARstart{N}{etwork} protocol is the set of communication rules between devices on the network, and it plays a crucial role in the whole network. Once there is a security vulnerability in the network protocol, it will lead to serious consequences. In 2014, OpenSSL, an open source implementation of the SSL (Secure Sockets Layer) security protocol, was exposed to a major security vulnerability called "HeartBleed", which led to the disclosure of confidential information such as user accounts, passwords, and private keys in the server.

\begin{figure}[!t]
    \centering
    \includegraphics[width=0.40\textwidth]{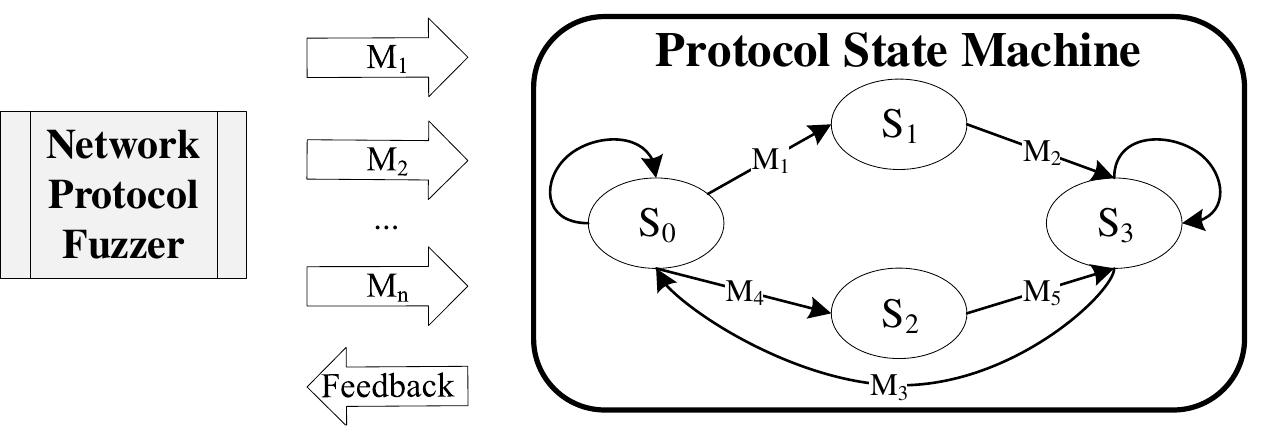}
    \caption{An example to explain the process of network protocol fuzzing.}
    \label{fig:explain-process}
\end{figure}

Coverage-based Greybox Fuzzing (CGF) is one of the most effective vulnerability discovery techniques, especially in fuzzing command-line inputs (AFL\cite{afl}, AFLFast\cite{bohme2017coverage}), library APIs (LibFuzzer\cite{libFuzzer}) and other stateless programs. However, it is very difficult for CGF to fuzz network protocol programs which have a huge state space. Only when an accurate input message sequence is defined can the state space of network protocol be traversed effectively. Network protocols can be modeled by finite state machine (FSM) theory. Specifically, considering a stateful protocol $P$ with $n$ states and $m$ state transitions. The state set $S$=<$S_0,S_1,...,S_n$>, the message sequence $M$=<$M_1,M_2,...,M_m$>, and the protocol state transition relationship can be expressed as a directed graph $G$=<$S, M$>. Figure \ref{fig:explain-process} gives an example of protocol state machine model. The model includes four states, and the initial state is $S_0$. In order to fuzz the $S_3$ of the protocol, $M_1$ and $M_2$ must be sent in order to reach $S_3$. However, most existing state-of-the-art greybox fuzzers such as AFL\citing{afl} and AFL++\citing{fioraldi2020afl++} do not consider the internal states of the protocol, the order and structure of the protocol messages, and thus only the initial state $S_0$ can be fuzzed. Therefore, traditional greybox fuzzing techniques are only suitable for fuzzing stateless programs.

Because of the problems mentioned above, the most widely used fuzzing technology for network protocol is still Stateful Blackbox Fuzzing (SBF) such as Peach\cite{peach} and BooFuzz\cite{boofuzz}. By constructing the protocol state machine model, SBF generates a large number of random or semi-random messages in different protocol states to find protocol bugs. However, SBF is a blackbox testing technology, which does not use program runtime information, such as coverage and data flow, to improve the quality of generated messages. Therefore, for SBF, a lot of time is often wasted in useless test cases and hard to cover deeper code coverage than CGF.

Recently, some researchers have applied CGF technique to network protocols called Stateful Coverage-based Greybox Fuzzing (SCGF) to solve the above problems. 
AFLNET\cite{aflnet} is the first greybox fuzzer for network protocol programs, which uses mutation-based method to generate messages. AFLNET uses the response codes as the states of the network protocol programs, which can accurately fuzz the actual states of the network protocol programs. Peach* \cite{9218603} and PAVFuzz \cite{PAVFuzz} use generation-based method to generate messages. Users describe the messages format and state transformation relationship of the protocol in the form of XML file after reading the protocol specification. Then Peach* and PAVFuzz can parse the XML format and transform it into the protocol state machine model to generate well-formatted messages. 

However, because of the complexity of stateful network protocol programs, the existing SCGF fuzzers are not efficient when fuzzing network protocols. Generally, the existing SCGF fuzzers face two challenges:

\textbf{Low speed.} Different from the traditional command-line inputs and library APIs programs, network protocol programs use socket communication to complete messages interaction. The communication speed of socket is slower than the above programs\cite{IPC-benchmark}. In this case, speeding up the fuzzing of network protocol programs can improve the efficiency of coverage and vulnerability discovery. The existing SCGF fuzzers first send a series of prefix messages, and then send mutated messages to fuzz the target state of network protocol programs. Specifically, if the state $S_3$ shown in Figure \ref{fig:explain-process} need to be fuzzed by a SCGF fuzzer, it sends prefix messages $M_1 $ and $M_2$ in turn first, and then sends mutated messages after reaching state $S_3$. Sometimes, a large number of prefix messages must be sent in order to reach some deep protocol states. More seriously, in the process of network protocol fuzz campaign, the prefix messages must be resent when the protocol states chosen to be fuzzed again. The above process results in a high time consumption and eventually leads to a low speed of network protocol fuzzing.

\textbf{Hard to cover deep states.} SCGF fuzzers are usually guided by the states of the network protocol programs. The basic idea of SCGF fuzzers is first select the target state as the fuzz object, and then select the message sequence that can reach the target state from the message sequence queue. In this case, it is likely that different message sequences can reach the same state in different ways. Taking Figure \ref{fig:explain-process} as an example, there are many ways to reach state $S_3$, such as $[S_0 \rightarrow S_1 \rightarrow S_3]$, $[S_0 \rightarrow S_2 \rightarrow S_3]$, $[S_0 \rightarrow S_1  \rightarrow S_3 \rightarrow S_0 \rightarrow S_2 \rightarrow S_3]$. Each additional message increases the time overhead of message sending and receiving. Therefore, most SCGF fuzzers prefer to the message sequence that can reach the target state faster (i.e. shorter prefix messages) to reduce messages sending and receiving time. However, the choice makes it difficult to cover the deep states and eventually deep paths and bugs can not be found.

In this paper, we propose SNPSFuzzer, a fast greybox fuzzer for stateful network protocol using snapshots. SNPSFuzzer dumps the context information of each state of the network protocol programs and restores it when the state needs to be fuzzed. Specifically, we save the process context (i.e. process snapshot) of the network protocol program to be fuzzed after SCGF fuzzers send prefix messages at the first time. Then whenever a specific state needs to be fuzzed, only the snapshot needs to be restored and the subsequent mutated messages are sent. This can save the time of sending prefix messages in each iteration of fuzzing and speed up the network protocols fuzzing. Moreover, in order to solve the problem that the existing SCGF fuzzers are difficult to cover deep states, we design a message chain analysis algorithm to explore more deeper protocol states.

In summary, the main contributions of this paper are as follows:
\begin{itemize}
\item We summarize the key ideas and processes of the existing SCGF fuzzers and investigate the reasons why the fuzzing speed of the existing SCGF fuzzers are slow. Based on the investigation, we propose to use snapshots technology to speed up network protocol fuzzing. 
\item We propose Snapshot Point Analysis (SPA) algorithm to help SCGF fuzzers to judge when to take and restore snapshots and Message Chain Analysis (MCA) algorithm to cover more and deeper protocol states. 
\item We implement a prototype of SNPSFuzzer and evaluate it on two widely used open source network protocol programs. The results show that compared with the state-of-the-art network protocol greybox fuzzer AFLNET, SNPSFuzzer can generally increase the speed of network protocol fuzzing by 112.0\%-168.9\% and improve path coverage by 21.4\%-27.5\%. It has also found an unknown vulnerability in a real network program.
\end{itemize}

\section{Background}

\subsection{Coverage-based Greybox Fuzzing}
    The Coverage-based Greybox Fuzzing (CGF) \cite{afl}, \cite{bohme2017coverage}, \cite{libFuzzer}, \cite{fioraldi2020afl++}, \cite{Honggfuzz}, \cite{cafl}, \cite{invscov}, \cite{RL-CGF} has received widespread attentions and achieved great success in finding vulnerabilities in command-line inputs, library APIs and other stateless programs. Unlike heavyweight technologies such as symbolic execution\cite{cadar2013symbolic} and taint analysis\cite{ganesh2009taint}, CGF is a fast and efficient technology, which uses lightweight instrument to obtain coverage information. Given an initial seed, CGF generates new test cases by mutation. If a new test case can trigger a new interesting path, it is retained; Otherwise, it will be discarded. The above process is executed during the fuzzing loop stage until timeout or termination signal is received.
    
    However, CGF can not fuzz stateful software such as network protocol programs. The network protocol programs are state-driven which can only be converted to a specific state after receiving a specific message. CGF cannot obtain the state information of network protocol programs, nor can it obtain the format and order of the messages.

\subsection{Stateful Coverage-based Greybox Fuzzing}
    \label{background:SCGF}
    Recent works have paid more attentions to the problem that traditional CGF can not fuzz network protocol programs\cite{Challenges}. In order to solve the problem, some researchers proposed Stateful Coverage-based Greybox Fuzzing (SCGF)\cite{aflnet},\cite{9218603},\cite{PAVFuzz}. SCGF adds the states and messages information of the network protocol programs to CGF and has the state awareness of network protocol programs. Its general idea is that in order to fuzz the specific state of network protocol programs, it first sends the prefix messages in turn to reach the specific state, and then sends the mutated messages in this state. In this process, \textit{message chain} and \textit{state chain} appears. 
    
    1) \textbf{Message chain} refers to the chain composed of messages sent successively in an iteration of fuzzing. A new message chain is generated during each iteration. 
    
    2) \textbf{State chain} refers to the chain composed of the states generated by network protocol programs in turn with the received message chain. At the beginning of each iteration, the network protocol program is in the initial state. Therefore, the first state of the state chain is the initial state $S_0$.

    \begin{figure*}[!t]
    \centering
    \quad
    \subfigure[Network protocol program]{
    \includegraphics[width=0.18\textwidth]{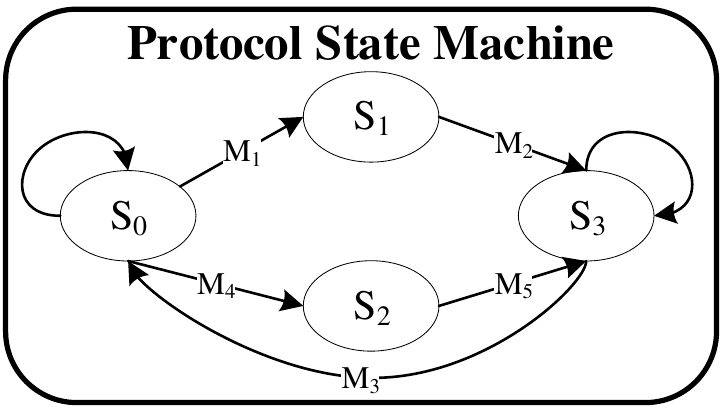}
    \label{example1-a.pdf}
    }
    \quad
    \subfigure[Message chain and state chain]{
    \includegraphics[width=0.72\textwidth]{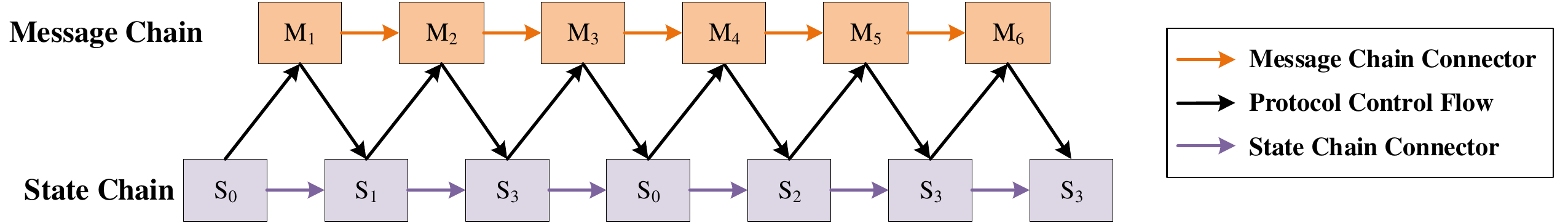}
    \label{example1-b.pdf}
    }
    \caption{An example to explain message chain and state chain.}
    \label{fig:example1}
    \end{figure*}

    Figure \ref{fig:example1} is an example to explain message chain and state chain. The state transition process of network protocol programs can be represented by Protocol State Machine (PSM) model, as shown in Figure  \ref{example1-a.pdf}. When the network protocol program starts at the beginning, it is in the initial state $S_0$. If message $M_1$ is received, the program is in state $S_1$; if message $M_4$ is received, the program is in state $S_2$; otherwise, it is still in state $S_0$. Suppose during an iteration of fuzzing, SCGF send messages $M_1,M_2,...,M_6$ successively, that is, message chain $MC$=$[M_1,M_2,M_3,M_4,M_5,M_6]$. At the beginning of the iteration, the network protocol program is in the initial state $S_0$. At this time of state $S_0$, the message $M_1$ is received, and the program will be in state $S_1$ according to the protocol state machine.  We can deduce the rest in the same manner. Finally the state chain of the network protocol program $SC$=$[S_0,S_1,S_3,S_0,S_2,S_3,S_3]$, as shown in Figure \ref{example1-b.pdf}.
    
\begin{figure}[!t]
    \centering
    \includegraphics[width=0.48\textwidth]{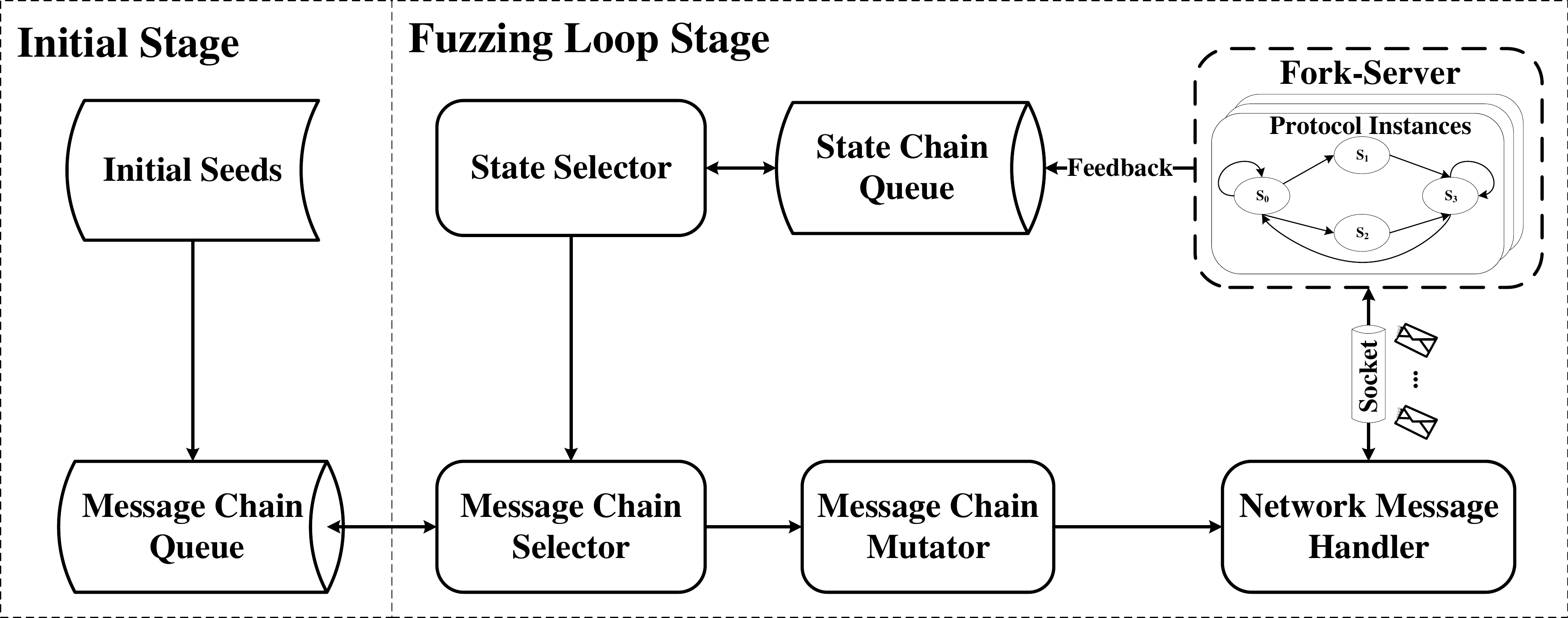}
    \caption{The general workflow of SCGF.}
    \label{fig:SCGF}
    \end{figure}

    
    Generally, the workflow of SCGF can be abstracted into two stages: \textit{initial stage} and \textit{fuzzing loop stage}, as shown in Figure \ref{fig:SCGF}. 
    
    \textbf{Initial stage}: One or more initial seeds provided by users for the network protocol program are added to the message chain queue. Similar to the seed queue of traditional CGF, the message chain queue maintains interesting message chain information. If a new message chain can find new code coverage or new states, SCGF regards it as interesting message chain. The initial stage is executed only once, and then enter into the fuzzing loop stage.
    
    \textbf{Fuzzing loop stage}: (1) Maintain a state chain queue that can be updated during fuzzing loop; (2) \textbf{State selector} selects the most interesting network protocol program state to be fuzzed as the target state from the state chain queue; (3) \textbf{Message chain selector} selects the message chain that can reach the target state from the message chain queue; (4) \textbf{Message chain mutator} mutates the specific messages in the message chain to generate new mutated messages; (5) \textbf{Network message handler} sends mutated messages to the network protocol program and receives the corresponding reply messages through socket communication to realize the state transition of the network protocol program; (6) Maintain the interesting message chain and state chain according to the feedback information such as states and code coverage; (7) Go back to Step (2).
    
\begin{figure}[!t]
    \centering
    \includegraphics[width=0.48\textwidth]{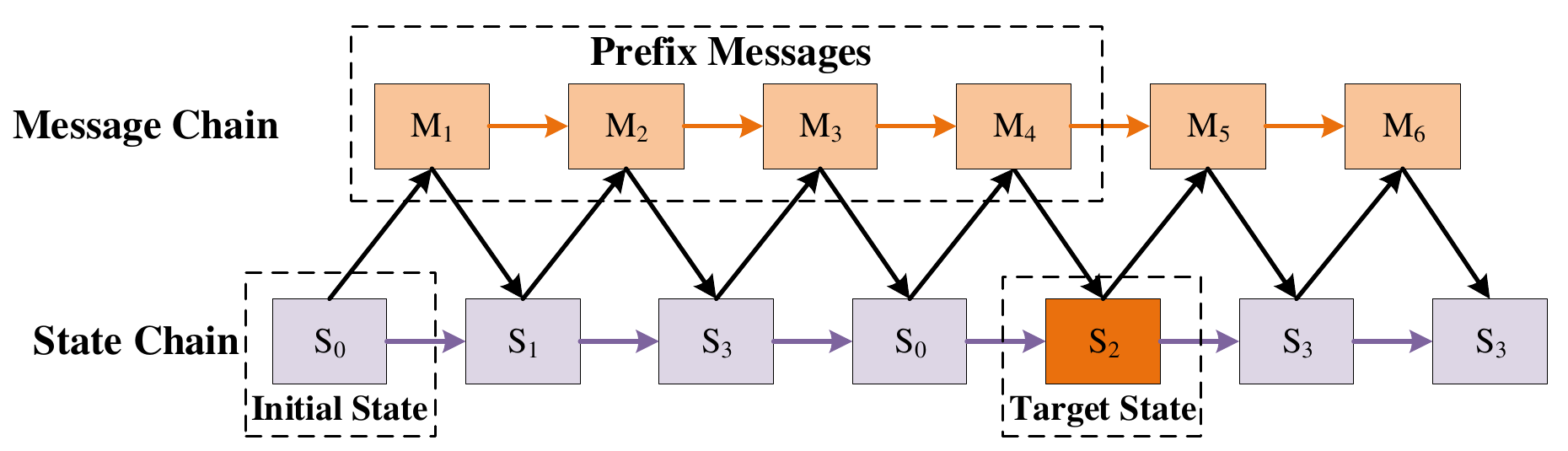}
    \caption{An example to explain prefix messages.}
    \label{fig:prefix_messages}
    \end{figure}

    SCGF saves the message chain and state chain information with the corresponding data structure to fuzz network protocol programs. In order to reach the target state of network protocol programs, SCGF sends corresponding messages in turn according to the data structure. Take Figure \ref{fig:prefix_messages} as an example, assuming that the state $S_2$ is the target state selected by the state selector. To start from the initial state $S_0$ to reach the state $S_2$, SCGF sends messages $M_1,...,M_4$ successively according to the information of $MC$ and $SC$, and then sends mutated messages to complete fuzzing state $S_2$. In the above process, the prefix messages (i.e. the messages sent successively to reach the target state) are $M_1,...,M_4$. Whenever the target state $S_2$ selected to be fuzzed, the prefix messages need to be resent. SCGF can complete fuzzing network protocol programs in different states. However, the speed of SCGF is very slow. Therefore, we aim at improving the speed of SCGF and finding more code coverage and vulnerabilities in the same time. 
    
\section{Motivation}

\subsection{The Speed of Stateful Coverage-based Greybox Fuzzing}  
\label{section:speed}
    Speed is very important for the efficiency of fuzzing. It turns out that even the most effective technique is less efficient than blackbox fuzzing if the time spent for generating a test case takes relatively too long\cite{7289448}. A typical example is symbolic execution and coverage-based greybox fuzzing. Symbolic execution uses program analysis and constraint solvers to generate high quality test cases, in which each test case can trigger a new path of the program. However, the process is very time-consuming. Different from symbolic execution, CGF uses the lightweight instrument to get code coverage without program analysis. Although the seed quality generated by CGF is not better than symbolic execution, the speed of CGF provides an important guarantee for its effectiveness. CGF has been proven to be effective and found a large number of real-world programs vulnerabilities\cite{li2018fuzzing}.
    
    To show the speed of SCGF, we conduct an experiment on a typical SCGF fuzzer AFLNET to evaluate the speed of fuzzing real-world network protocol programs. We select a total of five programs from \textit{ProFuzzBench} \cite{profuzzbench} (a state-of-art benchmark for stateful fuzzing of network protocols) in our experiment: Lightftp\cite{lightftp}, Live555\cite{live555}, Tinydtls\cite{tinydtls}, Kamailio\cite{kamailio}, Dcmtk\cite{dcmtk}. Because different instrument methods lead to different speeds, we all use \textbf{afl-clang-fast} instrument mode in this experiment. We calculate the execution speed according to Equation \eqref{speed}.
    
    \begin{equation}
    S=N/T \label{speed}
    \end{equation}
    where $N$ represents the total number of test cases executed, and $T$ represents the running time of fuzzer. 
    
    To better display the results, we use execution time per test case metric to represent the speed of SCGF. We calculate the excution time metric $E$ according to Equation \eqref{excution_time}.
    
    \begin{equation}
    E=1/S \label{excution_time}
    \end{equation}
    where $E$ is inversely proportional to $S$. A larger $S$ leads to a smaller $E$.
    
\begin{figure}[!t]
    \centering
    \includegraphics[width=0.48\textwidth]{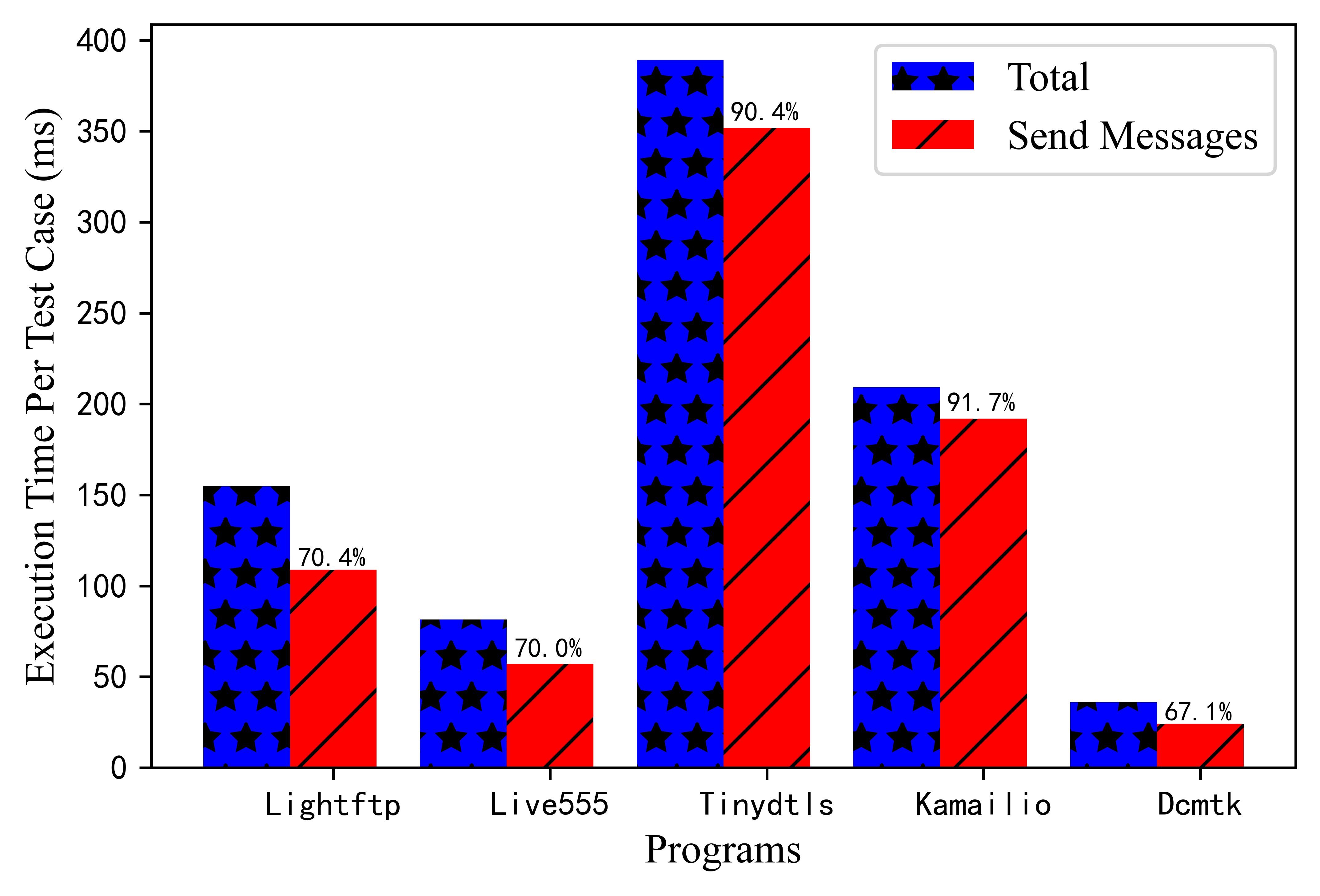}
    \caption{Execution time and messages sending time per test case.}
    \label{fig:execution-time}
    \end{figure}

    The results are shown in Figure \ref{fig:execution-time}. We find that for network protocol programs, the execution time of each test case is at the millisecond level, ranging from tens to hundreds of milliseconds. The average speed of AFLNET is at least two orders of magnitude slower than CGF fuzzers. This leads to the inefficiency of SCGF fuzzers represented by AFLNET. The results make us to think about the reason why network protocol fuzzing is slow. Therefore, we calculate the time spent for sending messages and show the percentage of messages sending time in the total execution time of each test case, as shown in Figure \ref{fig:execution-time}. We find that during the execution of each test case, most of the time was spent in the process of sending messages. Specifically, at least 67\% of the time is consumed in sending messages in the process of network protocol fuzzing. And the longer the execution time of each test case, the greater the percentage. This also shows that other time will not change significantly with different programs compared with the time for sending messages. 

\subsection{Further Analysis of Slow Speed and Motivation Description}

 
    As described in Section \ref{background:SCGF}, message chain refers to the messages sent successively in each iteration of network protocol fuzzing. Therefore, the messages sending time actually refers to the total time spent for sending each message in the message chain, including the time of sending messages and receiving reply messages. The messages sending time of each fuzzing iteration is positively correlated with the number of messages sent. Network protocol fuzzing is based on socket communication, which is very slow compared with command-line, library APIs fuzzing. The more messages are sent, the more socket communications are performed. Therefore, sending too many messages leads to a serious decline in the speed of network protocol fuzzing.
    
\begin{figure}[!t]
    \centering
    \includegraphics[width=0.48\textwidth]{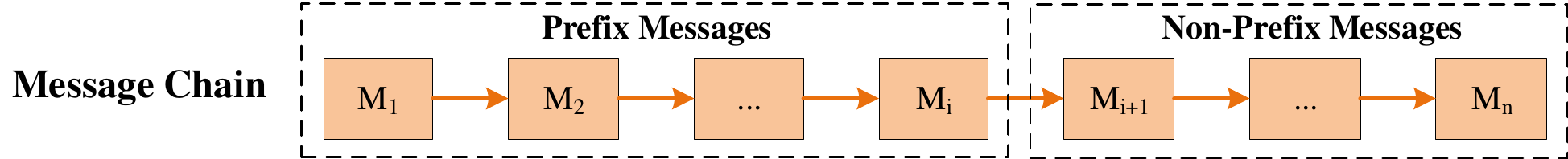}
    \caption{Division of message chain.}
    \label{fig:messages_Divide}
    \end{figure}

    On the other hand, network protocol programs are stateful. In order to fuzz the target state, prefix messages must be sent in advance. We can divide the messages sent each iteration into two parts: \textit{prefix messages} and \textit{non-prefix messages}, as shown in Figure \ref{fig:messages_Divide}. Suppose that during a fuzzing iteration, SCGF fuzzer sends a total of $n$ messages, where the first $i$ $(i < n)$ messages are prefix messages, and the number of non-prefix messages is $n-i$. Therefore, the more prefix messages sent, the less non-prefix messages need to be sent. 
    
   \textbf{Motivation 1.} SCGF first sends the corresponding prefix messages to reach the target state, and then sends mutated messages in this state to fuzz the target state of the network protocol program. If the network protocol program in the target state can be saved, we do not need to send prefix messages in the process of fuzzing the target state. This can save the time cost of sending prefix messages and speed up the network protocol fuzzing. Moreover, we can save more messages sending time overhead to speed up the network protocol fuzzing if the length of prefix messages is longer.
    
\begin{figure}[!t]
    \centering
    \includegraphics[width=0.48\textwidth]{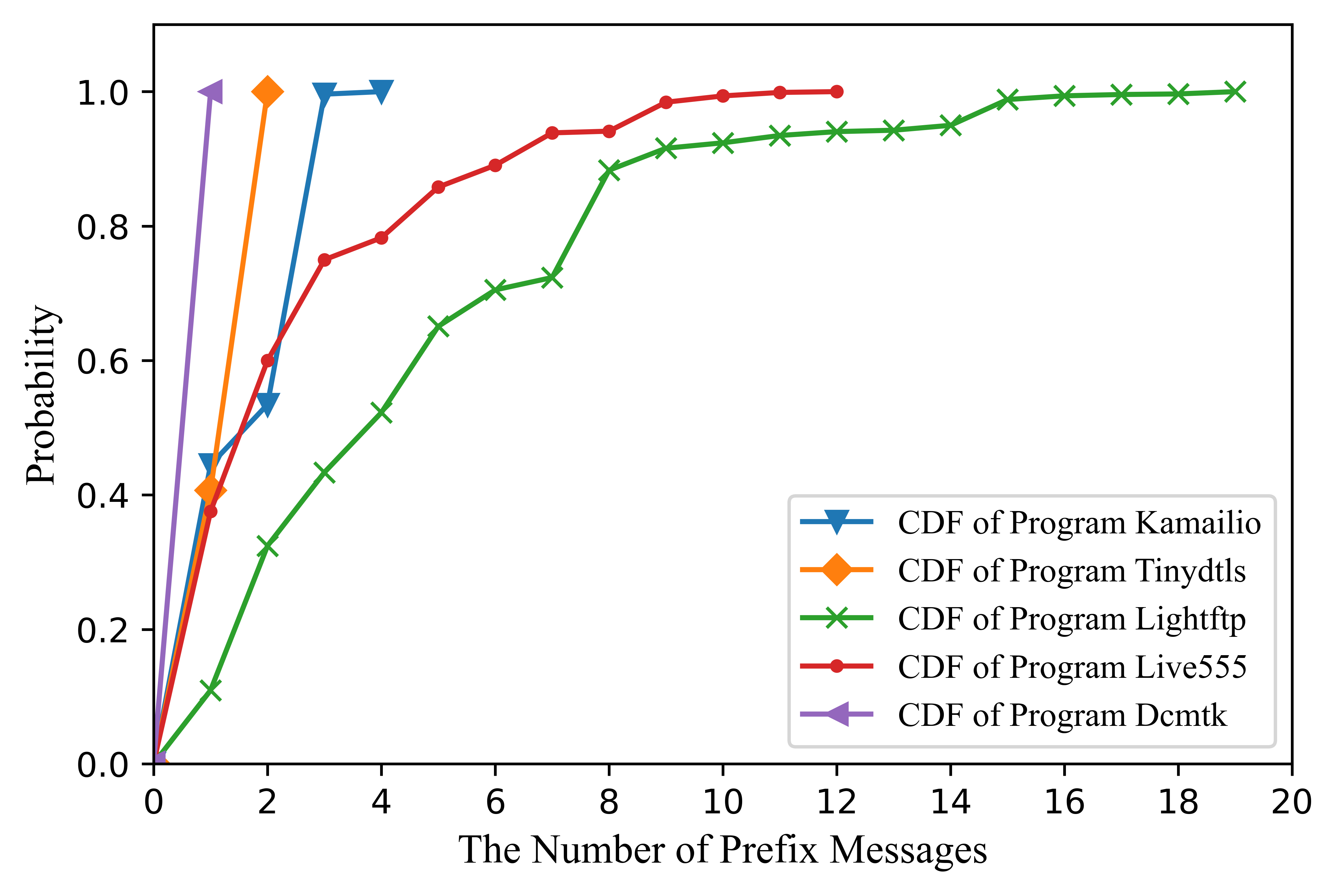}
    \caption{CDF for the number of prefix messages of five typical programs.}
    \label{fig:CDF_ALL}
    \end{figure}

    According to the Motivation 1, we further analyze the Cumulative Distribution Function (CDF) of average number of prefix messages when AFLNET fuzzs different network protocol programs. The results are shown in Figure \ref{fig:CDF_ALL}. We find that the average number of prefix messages is different for different protocols. For program Live555 and Lightftp, the numbers of prefix messages are relatively bigger. However, for program Kamailio, Tinydtls and Dcmtk, the numbers of prefix messages are less than four. 
    
    \textbf{Motivation 2.} Different number of prefix messages will affect the efficiency of network protocol fuzzing in this situation. Specifically, different number of prefix messages represents different state depths of reaching the target state. Taking Figure \ref{fig:explain-process} as an example, there are multiple ways to reach the state $S_3$, such as $[S_0 \rightarrow S_1 \rightarrow S_3]$, $[S_0 \rightarrow S_2 \rightarrow S_3]$, $[S_0 \rightarrow S_1 \rightarrow S_3 \rightarrow S_0 \rightarrow S_2 \rightarrow S_3]$. Each additional message increases the time overhead of message sending and receiving. Therefore, most SCGF fuzzers prefer to the message chain that can reach the target state faster (i.e. the shorter prefix messages) to reduce the time of messages sending and receiving. However, the choice makes it difficult to cover the deep states and eventually deep paths and bugs can not be found. On the basis of the snapshots technology, increasing the number of prefix messages does not increase the time overhead of messages sending and receiving. Therefore, we design an algorithm to increase the number of prefix messages to explore more and deeper states.

\section{Overview of SNPSFuzzer}

\begin{figure*}[!t]
    \centering
    \includegraphics[width=0.78\textwidth]{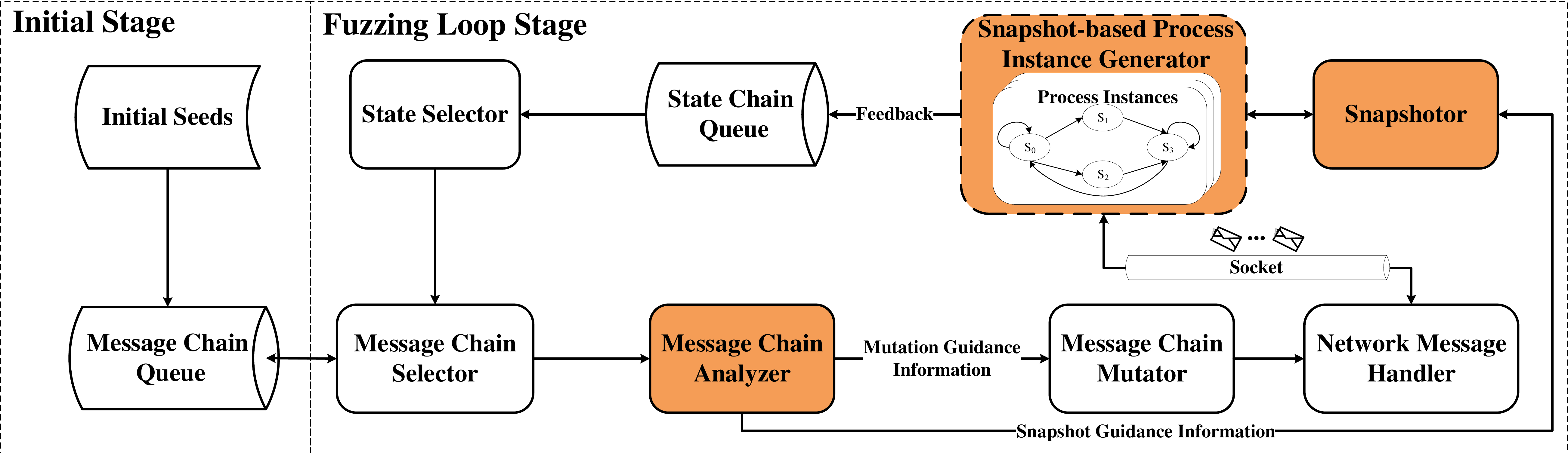}
    \caption{Overview of SNPSFuzzer.}
    \label{fig:snpsfuzzer-overview}
\end{figure*}

\subsection{Framework}
    The basic workflow and main components of SNPSFuzzer are shown in Figure \ref{fig:snpsfuzzer-overview}. The basic workflow is divided into two stages: \textit{initial stage} and \textit{fuzzing loop stage}.

    At the beginning of a fuzz campaign, SNPSFuzzer is in the initial stage. Given one or more initial seeds provided by users for the network protocol program, SNPSFuzzer adds them to the message chain queue as initial queue. Similar to the seed queue of traditional CGF, the message chain queue maintains interesting message chain information. If the message chain can find new code coverage or states, SNPSFuzzer regards it as interesting message chain.
    
    In the fuzzing loop stage, the following steps are carried out in turn:
    
    (1) Maintain a state chain queue that can be updated during fuzzing loop;
    
    (2) State selector selects the most interesting network protocol program state to be fuzzed as target state from the state chain queue;
    
    (3) Message chain selector selects the message chain that can reach the target state from the message chain queue;
    
    (4) \textbf{Message chain analyzer} analyzes the message chain selected by the message chain selector to determine: i) whether to take and restore snapshots of the Protocol Under Test (PUT);  ii) the most interesting parts of message chain. For the former, if the PUT needs to be taken snapshots, the message chain analyzer will pass the snapshot guidance information to the \textbf{snapshotor} to $checkpoint$ the process instance generated by the snapshot-based process instance generator; and if the PUT already has a snapshot, the message chain analyzer calls the snapshotor to $restore$ the snapshot. For the latter, the message chain analyzer passes the mutation guidance information to the message chain mutator to mutate the most interesting parts of message chain;
    
    (5) Message chain mutator mutates the specific messages in the message chain to generate new mutated messages;
    
    (6) Network message handler sends mutated messages to the PUT generated by the \textbf{snapshot-based process instance generator} and receives the corresponding reply messages through socket communication to realize the state transition of the PUT. The Snapshot-based process instance generator has two generation modes: the traditional Fork-Server mode and the snapshot-based process instance generation mode. When no snapshot is needed in fuzzing PUTs, the traditional Fork-Server mode is used to generate process instances. Otherwise, the snapshot-based process instance generation mode is used;
    
    (7) Maintain interesting message chain and state chain according to the feedback information;
    
    (8) Go back to Step (2).
    
    Based on the existing SCGF fuzzer, we mainly modify it in the fuzzing loop stage and add three main components: \textit{snapshot-based process instance generator}, \textit{snapshotor} and \textit{message chain analyzer}. These three components will be described in detail later in this section.

\subsection{Snapshot-based Process Instance Generator}
\label{section:instance generator}
    Among the existing snapshot technologies, one of the most widely used is the virtual machine snapshot technology\cite{cui2013hotsnap},\cite{cui2018snapfiner},\cite{raju2019snaps} which can save the state and data information of the virtual machine at a specific time. However, this technology has large granularity, and the information taken and restored is too large and redundant. Furthermore, the speed of virtual machine snapshot technology is slow. Therefore, it is not suitable for the fast fuzzing scenario. Another way is to save some key information of a program such as the value of program memory. In this situation, we need to have a good understanding of the internal implementation details of a program to save the most key program information and overwrite the original value at a specific time. However, the implementation details and internal states of different network programs are often different. It is difficult to judge what key information different programs should save. Once the key information of programs is not designed to be saved, it can not restore the correct program context and even cause the programs to crash directly. There will be a large number of false positives in the fuzzing results.  Therefore, it is necessary to consider a snapshot technology with appropriate granularity. It can not only quickly complete taking and restoring snapshots to speed up the network protocol fuzzing, but also be applied to various network programs accurately and effectively.

    \begin{figure}[!t]
    \centering
    \includegraphics[width=0.48\textwidth]{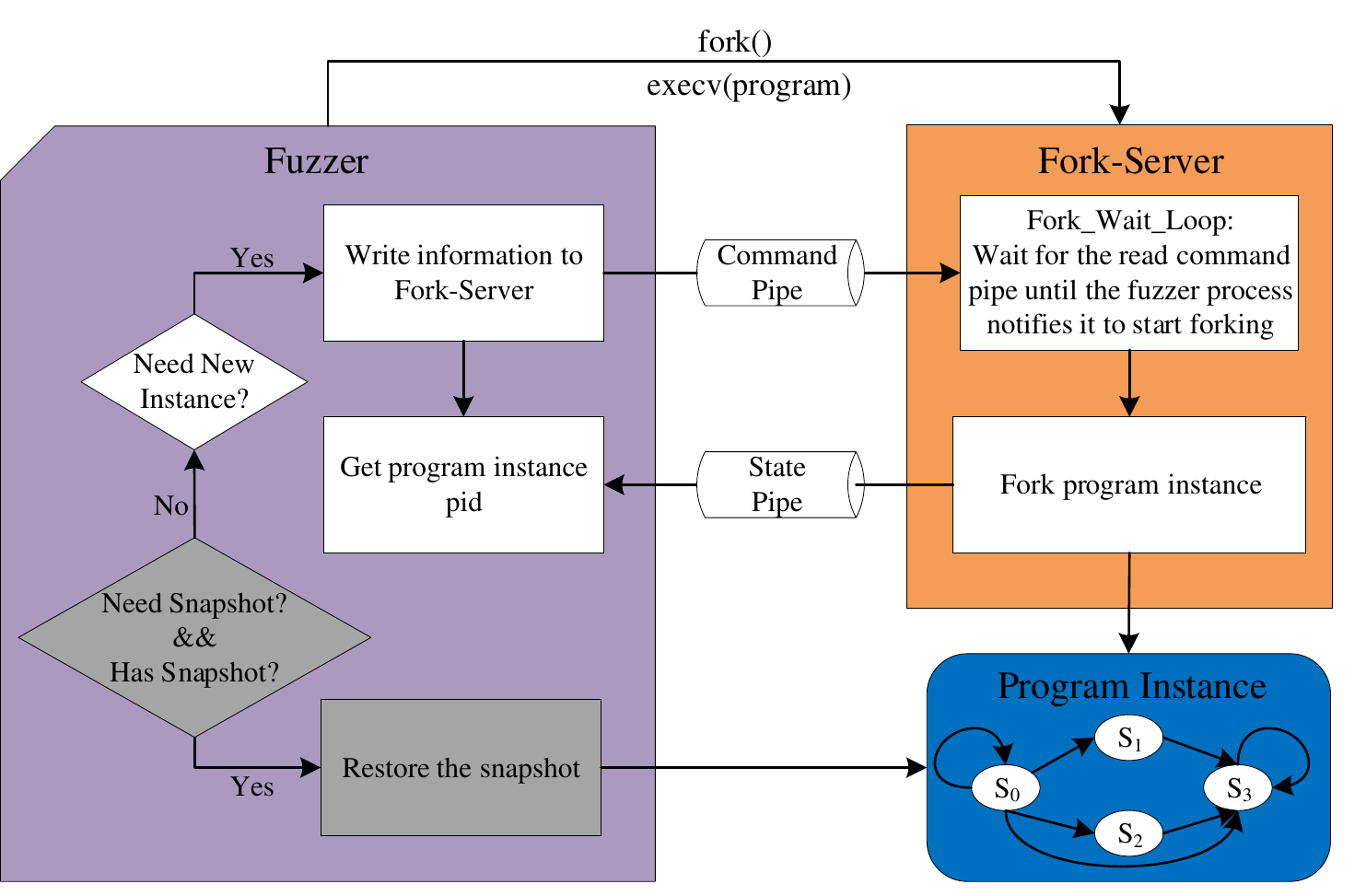}
    \caption{The workflow of the snapshot-based process instance generator.}
    \label{fig:instance generator}
    \end{figure}

    We find that there are two processes in a fuzz campaign: \textit{fuzzer process} and \textit{PUT process}. No matter what the PUT is fuzzed, the program exists in the form of process universally. As the basic unit of system resource allocation and scheduling, process has independent code and data space. Generally, it does not interfere with other processes in the process of taking and restoring snapshots. At the same time, the speed of the process level snapshot is reasonable and it is faster than the network messages sending process. Therefore, we use the process level as the granularity of snapshot technology.

    In order to realize the program process level snapshot technology analysis, we need to generate process level instances. Figure \ref{fig:instance generator} shows the workflow of the snapshot-based process instance generator. There are three process instances in the entire workflow: \textit{fuzzer process}, \textit{Fork-Server process} and \textit{program instance process}. Among them, the Fork-Server process is generated by the fuzzer process through $fork()$ and $execv()$, and they communicate through command pipe and status pipe. To generate process instances, the fuzzer process first judges whether an iteration of network protocol fuzzing meets two conditions: 1) whether to use snapshots; 2) whether to has snapshots. If both conditions are satisfied, the snapshotor use $restore$ function to obtain the process instance of the previous snapshots. Otherwise, the Fork-Server process is informed through the pipe to generate a new process instance and returns the corresponding process $pid$.

\subsection{Snapshotor}
    Snapshotor is an important component for taking and restoring snapshots. We use the program process level as the granularity of snapshots according to the analysis in Section \ref{section:instance generator}. Therefore, the snapshotor is dedicated to taking and restoring snapshots of program process. 
    
    In order to better realize the snapshots of PUTs, the snapshotor must support the complete preservation and restoration of process context information for PUTs. In the scenario of network protocol program fuzzing, the context information of the PUT has three kinds of features: 1) Features of the PUT running environment; 2) Features of the PUT itself; 3) New features of the PUT brought by the instrument of fuzzers.
    
    \textbf{Features of the PUT running environment:} At present, most open source PUTs and fuzzers run in the Linux OS. Moreover, we want to fuzz the new version of the PUT as much as possible. Therefore, we need to run them in a newer Linux kernel version. 
    
    \textbf{Features of the PUT itself:} Different from other programs, the PUT has two characteristics: 1) The PUT uses the network programming and needs the $socket$ interface; 2) The PUT is often the multi-process or multi-thread program. Whenever a legal client request is received, the PUT starts up a new process or thread to handle the request and achieve the goal of handling multiple requests at the same time.
    
    \textbf{New features of the PUT brought by the instrument of fuzzers:} The greybox fuzzer obtains the code coverage information of the program through instrument and uses the information to determine whether interesting seeds are generated. In this workflow, the fuzzer process needs to interact with the program instance process, so as to realize the sharing of code coverage information. IPC (Inter-Process Communication) is the main method to meet the above requirements. There are many ways in IPC, such as message queue, pipe, shared memory. And one of the state-of-the-art greybox fuzzer AFL uses shared memory to share code coverage information. 
    
    We hope that the snapshotor can support the the above three features of the PUT. Therefore, we investigate the existing process snapshotors and obtain the results shown in Table \ref{tab:process_snapshotors}\cite{snapshotor-comparasion}. We find that CRIU meets all features of the PUT, so we choose CRIU as the snapshotor.
    
\begin{table}[htbp]
\centering
\caption{Comparison of different process snapshotors}
\label{tab:process_snapshotors}
\resizebox{0.48\textwidth}{!}{%
\begin{tabular}{|c|c|c|c|c|c|}
\hline
\multicolumn{2}{|c|}{Features}                             & \multicolumn{1}{c|}{CRIU} & DMTCP & \multicolumn{1}{c|}{BLCR} & OpenVZ \\ \hline
Linux Kernel                                   & 3.11 or later  & \checkmark    & \checkmark   & \checkmark                       & \XSolidBrush      \\ \hline
\multirow{3}{*}{Sockets}          & Unix Scokets    & \checkmark                       & \checkmark   & \XSolidBrush                        & \checkmark    \\ \cline{2-6}       & UDP Sockets     & \checkmark                       & \XSolidBrush     & \XSolidBrush                         & \checkmark    \\ \cline{2-6}
        & TCP Sockets     & \checkmark                       & \checkmark   & \XSolidBrush                         & \checkmark    \\ \hline
\multirow{2}{*}{Multiple Thread/Process} & Multithread     & \checkmark   & \checkmark   & \checkmark                & \checkmark    \\ \cline{2-6} 
 & Multiprocess    & \checkmark    & \checkmark   & \checkmark       & \checkmark    \\ \hline
\multirow{3}{*}{Shared Resources}        & Pipes           & \checkmark    & \checkmark   & \XSolidBrush                         & \checkmark    \\ \cline{2-6}   & System V IPC    & \checkmark   & \checkmark   & \XSolidBrush          & \checkmark    \\ \cline{2-6} 
 & Memory Mappings & \checkmark      & \checkmark   & \XSolidBrush   & \checkmark    \\ \hline
\end{tabular}%
}
\end{table}
    
    The snapshotor CRIU realizes two functions: $checkpoint$ and $restore$. The specific processes are as follows:  
    
    \textbf{Checkpoint.} Take the process snapshot, including three steps: 1) Freeze the process and its child processes to avoid errors caused by process execution during the snapshot taking; 2) Save the complete status information of the process to the image file; 3) Kill the program instance process.
    
    \textbf{Restore.} Restore the process snapshot, including five steps: 1) Read the image file and obtain the snapshot information; 2) Clone process and its child processes; 3) Restore the basic resources for the process; 4) Jump to the state when the process takes a snapshot; 5) Resume the remaining process tasks.
    
    \subsection{Message Chain Analyzer}
    \textbf{Basic ideas:} The location selection of snapshot taking and restoring will affect the efficiency of network protocol fuzzing. If the location selection is not good, it will not accelerate the speed of fuzzing and even affects the accuracy of the fuzzing results. The existing stateful coverage-based greybox fuzzing (SCGF) adds the state information of the network protocol programs to CGF, and has the state awareness of the network protocol programs. Its general idea is that in order to fuzz the target state of the PUT, SCGF fuzzers first send the corresponding prefix messages in turn to reach the target state, and then send mutated messages in the target state. Whenever the target state needs to be fuzzed, the above prefix messages need to be resent. The speed of messages sending is slow, which leads to the low efficiency of SCGF.
    
\begin{figure}[!t]
    \centering
    \includegraphics[width=0.48\textwidth]{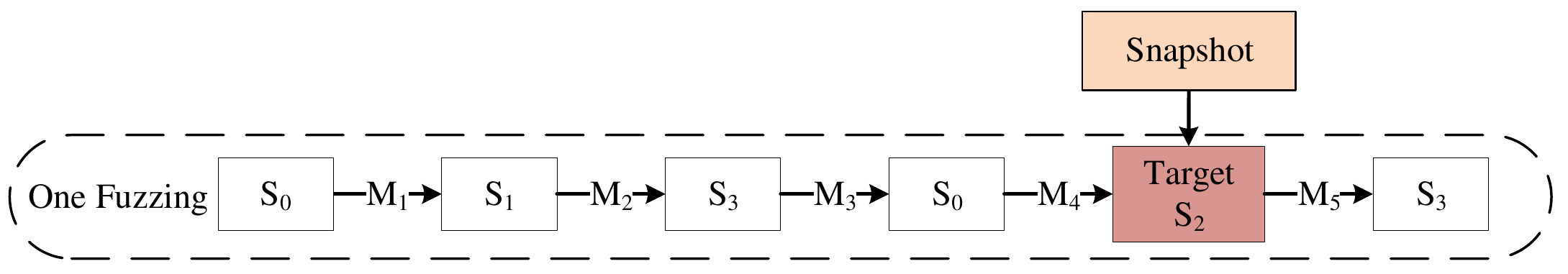}
    \caption{Basic ideas of message chain analyzer.}
    \label{fig:basic-idea}
    \end{figure}

    Based on the above observations, we propose to use snapshot technology to save the state of the network protocol program process (i.e. the program process context information) at a certain time and restore the state snapshot when the state needs to be fuzzed later. The method can avoid the time overhead of repeatedly sending prefix messages. Take Figure \ref{fig:example1} as an example, when the network program is in state $S_2$, we use snapshot technology to save the state of the program. Whenever state $S_2$ needs to be fuzzed, we restore the network program snapshot in state $S_2$, as shown in Figure \ref{fig:basic-idea}. It can reduce the overhead of sending prefix messages $M_1,..., M_4$ and thus speed up the network protocol program fuzzing.
    
    \textbf{Details:} We propose the Snapshot Point Analysis (SPA) algorithm to judge when to take and restore snapshots, as shown in Algorithm \ref{alg:algorithm_SPA}. 
    Whether to take a snapshot is related to the length of prefix messages in the message chain. If the target state is the initial state $S_0$, the length of the prefix messages is 0. In other words, the state can be reached without sending prefix messages. For this case, we do not need to take a snapshot; Otherwise, take a snapshot. Specifically, the entire SPA algorithm can be divided into the following three steps:
    
    1) Algorithm SPA has three input parameters: message chain $MC$, state chain $SC$ and target state $State$. These three parameters will be called by function $SPLIT\_MC$ to divide the message chain $MC$ into three parts $MP$, $MI$ and $MS$ (Line 4), where $MP$ represents the prefix messages to reach the target state; $MI$ represents the infix messages when the PUT is in the target state; $MS$ represents the suffix messages remaining in the message chain. We do this for two purposes, as shown below:
    \begin{itemize}
    \item The traditional CGF fuzzer is not aware of the network program states and can not identify which message will be more effective. Therefore, it treats all messages in the message chain equally and mutate the the entire message chain. Specifically, given a message chain $MC$ with $n$ $(n>0)$ messages, that is, $MC$ = [$M_1, .., M_n$]. Assuming that the message $M_i$ $(0<i<n)$ is the most interesting message, CGF does not focus on $M_i$ and repeatedly mutates the uninterested messages $M_1$ to $M_{i-1}$ before mutating the message $M_i$. By splitting the $MC$, we can effectively mutate the most interesting messages to improve the efficiency of network protocol fuzzing (Line 5-Line 6); 
    \item It is more effective to determine when a snapshot should be taken. As described in Section \ref{background:SCGF}, the target state is the most interesting state in the current fuzzing iteration. After sending the prefix messages, the program is in the target state. Exploring it is more likely to find more code coverage and vulnerabilities. Therefore, we judge whether a snapshot can be taken according to the length of prefix messages to quickly fuzz the most interesting state. 
    \end{itemize}

    \begin{algorithm}
     \caption{Snapshot Point Analysis (SPA)}
     \label{alg:algorithm_SPA}
     \begin{algorithmic}[1]
      \Function{SPA}{$MC, SC, State$}
      \State{$can\_snapshot = \textbf{FALSE}$}
      \State{$has\_snapshot = \textbf{FALSE}$}
      \State{$MP, MI, MS = SPLIT\_MC(MC, SC, State)$}
      
      \State{$MI^\prime = MUTATE(MI)$}
      \State{$MC^\prime = <MP, MI^\prime, MS>$}
      \If{LENGTH($MP$) > 0} 
        \State{$can\_snapshot = \textbf{TRUE}$}
      \EndIf
      \If{$can\_snapshot$}
        \If{$has\_snapshot$}
            \State{RESTORE\_SNAPSHOT()}
        \Else
            \State{SEND\_MESSAGES($MP$)}
            \State{CHECKPOINT\_SNAPSHOT()}
            \State{$has\_snapshot = \textbf{TRUE}$}
            \State{RESTORE\_SNAPSHOT()}
        \EndIf
        \State{SEND\_MESSAGES($MI^\prime$)}
        \State{SEND\_MESSAGES($MS$)}
    \Else
        \State{SEND\_MESSAGES($MC^\prime$)}
      \EndIf
      \EndFunction
     \end{algorithmic}
    \end{algorithm}

    We design a message chain analysis algorithm to divide the message chain, as shown in Algorithm \ref{alg:algorithm_MCA}. When the infix messages are determined, the prefix messages and suffix messages in the message chain will also be determined. Therefore, we define two variables $MI\_Start$ and $MI\_Count$ to represent the starting position and number of infix messages respectively. We infer from the end of the state chain until the location of the target state appears, and update $MI\_Start$ to the location. Then, we continue to judge the number of consecutive occurrences of the target state to update $MI\_Start$ and $MI\_Count$. Finally, we use the two variables to split the message chain. 
    
\begin{figure*}[!t]
    \centering
    \quad
    \subfigure[The target state occurs once in the state chain]{
    \includegraphics[width=0.46\textwidth]{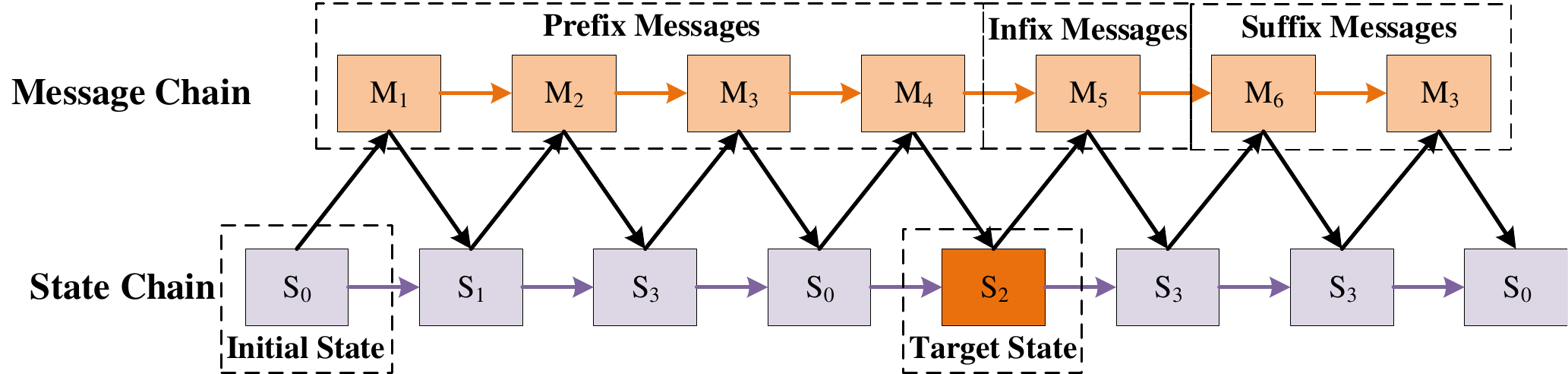}
    \label{example3-a}
    }
    \quad
    \subfigure[The target state occurs multiple times in the state chain]{
    \includegraphics[width=0.46\textwidth]{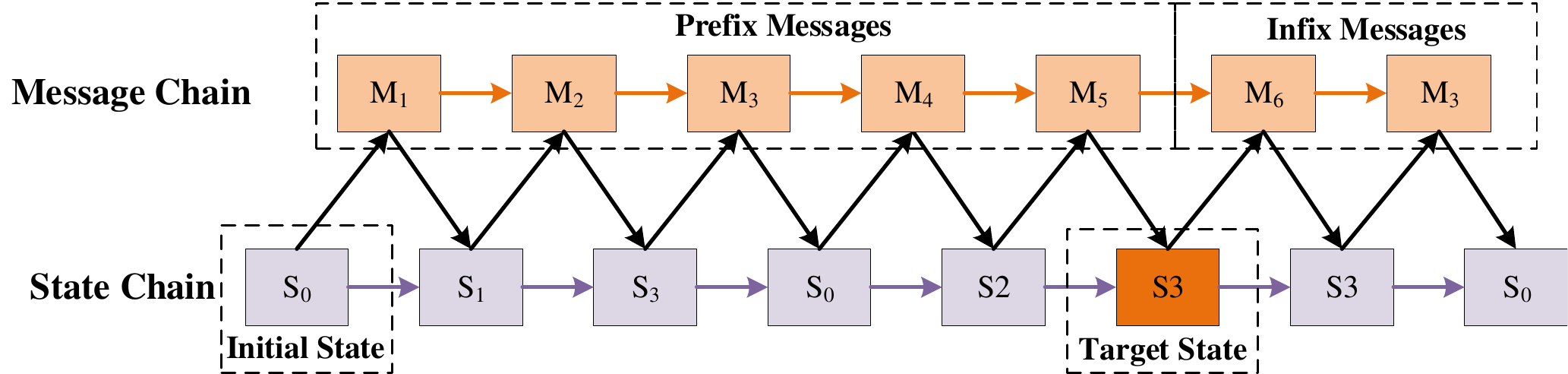}
    \label{example3-b}
    }
    \caption{Examples to explain function SPLIT\_MC.}
    \label{fig:example3}
    \end{figure*}

    Take the Figure \ref{fig:example3} as examples to illustrate the idea of function SPLIT\_MC. Figure \ref{example3-a} shows the situation when the target state occurs once in the state chain. Assuming that the selected target state is state $S_2$ in an iteration, Algorithm MCA sets the messages $[M_1, M_2, M_3, M_4]$ to reach the state $S_2$ as prefix messages, the messages sent $[M_5]$ when the PUT is in the state $S_2$ as infix messages, and the remaining messages $[M_6, M_3]$ as suffix messages.

    Figure \ref{example3-b} shows a case that the target state occurs multiple times in the state chain. Assuming that the selected target state is state $S_3$ in an iteration, we can find that the state $S_3$ appears three times in the state chain $SC$. Therefore, there are three choices in this case, and different choices lead to different efficiencies of the final results. The existing SCGF fuzzers often select the target state $S_3$ appears for the first time in the state chain as the target point, which can reach the target state faster and reduce messages sending and receiving time. However, this choice makes it difficult to cover the deep states and eventually deep paths and bugs will not be found. On the basis of the snapshots technology, increasing the number of prefix messages does not increase the time overhead of messages sending and receiving. Because once we select the target state, we take snapshot for the PUT in this state. Whenever the state is fuzzed, we can directly restore the state snapshot. In the entire process, there is no need to send  messages $MP$ again. Therefore, we select the last consecutive target state $S_3$ in the state chain as the target point to avoid the time overhead caused by the prefix messages and explore more and deeper states of the PUT. We will compare the performance impact of these two options in Section \ref{evalution-chapter}.

    \begin{algorithm}[t]
     \caption{Message Chain Analysis (MCA)}
     \label{alg:algorithm_MCA}
     \begin{algorithmic}[1]
      \Function{SPLIT\_MC}{$MC, SC, State$}
      \State{$MI\_Start = 0$}
      \State{$MI\_Count = 0$}
        \For{$i$ from LENGTH($SC$)-1 down to 1}
        \If{$SC[i] == State$}
            \State{$MI\_Start = i$}
            
            \For{$j$ from $i-1$ down to 1}
            \If{$SC[j]$ == $State$}
                \State{$MI\_Count$ += $1$}
                \State{$MI\_Start = j$}
            \Else
                \State{\textbf{break}}
            \EndIf
            \EndFor
             
            \State{\textbf{break}}
        \EndIf
        \EndFor

        \State{$MP$=$MC$[1:MI\_Start-1]}
        \State{$MI$=$MC$[MI\_Start:MI\_Start+MI\_Count]}
        \State{$MS$=$MC$[MI\_Start+MI\_Count+1: LENGTH($MC$)]}
      \EndFunction 
      
     \end{algorithmic}
    \end{algorithm}
    
    2) Algorithm SPA maintains two important bool variables: $can\_ Snapshot$ and $has\_Snapshot$, where the former represents whether a snapshot can be taken under a certain state of the PUT, the latter represents whether there is a corresponding snapshot in this state. We set them to FALSE during the initialization stage (Line 2-Line 3). After using function SPLIT\_MC to split the $MC$, we can determine whether the snapshot can be taken by judging whether $MP$ exists (i.e. whether the length of $MP$ is greater than 0) (Line 7). If the snapshot can not be taken, the message chain $MC ^\prime$ will be sent directly (Line 22). The reason is that when the length of $MP$ is 0, the target state is the initial state which can be reached without sending prefix messages. The snapshot taking and restoring time is usually shorter than the time of sending prefix messages. Once a snapshot is taken, there is no need to take snapshot in the next iteration of fuzzing the target state repeatedly and just restore the snapshot. Therefore, using snapshot can speed up the network protocol fuzzing when prefix messages $MP$ exist.
    
    3) After determining whether snapshot can be taken in the target state, we need to further judge whether there is a snapshot in this state. If the snapshot exists, function RESTORE\_SNAPSHOT is directly called to restore snapshots (Line 11-Line 12). Otherwise, it first calls function SEND\_MESSAGES to send messages $MP$ and receive the corresponding reply messages, so that the PUT reaches the target state. Then it calls function CHECKPOINT\_SNAPSHOT to take a snapshot of the PUT and set the snapshot flag $has\_snapshot$ to TRUE. Finally, restore the snapshot of the PUT (i.e. the PUT in the target state) (Line 13-Line 17).
    
    We send mutated messages $MI^\prime$ (Line 19) and suffix messages $MS$ (Line 20) in turn to complete fuzzing the target state after restoring the snapshot. It is note that the Algorithm SPA needs to take snapshot only once when the target state is selected in each iteration of fuzzing. In the subsequent fuzzing process, we restore the snapshot for reducing the time overhead of repeatedly sending prefix messages.

\section{Evaluation}
\label{evalution-chapter}
    \textbf{Implementation.} We implemented SNPSFuzzer based on the widely used SCGF fuzzer AFLNET\cite{aflnet}. We use snapshotor CRIU\cite{criu} in our implementation. The snapshot-based process instance generator and message chain analyzer are mainly implemented in the fuzzing loop stage used in C language.
    
    \textbf{Experimental setup.} We selected two famous stateful network protocols, DTLS\cite{dtls} and SIP\cite{sip}, as evaluation programs. These programs have been widely used at present. Therefore, we believe that these two programs are representative. Specifically, we chose open-source Tinydtls\cite{tinydtls} and Kamailio\cite{kamailio} as the concrete implementations corresponding to these two protocols. 

    \begin{table}[htbp]
    \centering
    \caption{Program's description and size in kilo of lines of code}
    \label{tab:program_description}
    \resizebox{0.485\textwidth}{!}{%
    \begin{tabular}{llll}
    \hline
    Program  & Size & Protocol & Description   \\ \hline
    Tinydtls & 11.4K          & DTLS      &   DTLS server implementation for IoT devices      \\ \hline
    Kamailio & 1074.0K        & SIP       & SIP signaling server implementation for large deployments \\ \hline

    \end{tabular}%
    }
    \end{table}
    
    We compared SNPSFuzzer with two baseline open-source network protocol greybox fuzzers: AFLNET\cite{aflnet} and AFLNWE\cite{aflnwe}. AFLNET is the first greybox fuzzer for network protocol programs, which adopts mutation-based method to generate messages. It uses the response codes as the states of the network protocol programs, which can accurately fuzz the actual states of the network protocol programs. AFLNWE is a stateless coverage-guided fuzzer. It changes the file input mode of AFL into the network protocol socket sending mode. At present, there are other SCGF fuzzers, such as Peach*\cite{9218603} and PAVFuzz\cite{PAVFuzz}, but they are not open source. We hope to make further comparison with them after they open source. Meanwhile, we call the situation without using Algorithm MCA (Message Chain Analysis) as \textbf{SNPSFuzzer-tmp} mode and this mode is also added to the comparison object to evaluate the performance of Algorithm MCA. 
    
    All our experiments are run on a server of Intel Xeon Gold 6230R with 104 CPU cores and 125GB RAM, which runs Ubuntu 16.04 LTS. In order to ensure the fairness of our experiments, we provided the same initial seeds for different fuzzers. At the same time, we conducted each experiment for 24 hours and repeated each experiment 5 times to determine the statistical significance of the results.
    
    \textbf{Research Questions.} By comparing with two network protocol greybox fuzzers AFLNET and AFLNWE, we answer the following research questions:
    
    \textbf{RQ1.} How do the results of execution speed compare with the state-of-the-art stateful coverage-based greybox fuzzer AFLNET?
    
    \textbf{RQ2.} How many messages can SNPSFuzzer save compared with the total number of messages sent by AFLNET?
    
    \textbf{RQ3.} How do the results of code coverage in network protocol fuzzing compare to the previous network protocol greybox fuzzers?
    
    \textbf{RQ4.} How do the results of vulnerability discovery in network protocol fuzzing compare to the previous network protocol greybox fuzzers?
    
\subsection{Speed Evaluation (RQ1)}
    SNPSFuzzer uses snapshot technology to save the state of a specific network protocol program. Whenever it is necessary to fuzz the state, SNPSFuzzer does not need to send prefix messages frequently. Therefore, we can save the time of frequently sending prefix messages and thus speed up the network protocols fuzzing.
    
    Because SNPSFuzzer is expanded on the basis of AFLNET, we mainly compared the execution speed with AFLNET. Among them, the speed of network protocol fuzzing is calculated according to the Equation \eqref{speed} in Section \ref{section:speed}. We record fuzzing speed of different fuzzers in the same 24 hours. The results are shown in Table \ref{tab:experiment_speed}, where $R_T$ and $R_S$ represent the improvement of SNPSFuzzer-tmp and SNPSFuzzer compared with AFLNET respectively. 
    
    \textbf{Answer to RQ1.} We find that both SNPSFuzzer-tmp and SNPSFuzzer can speed up the network protocol fuzzing. Specifically, the speed of SNPSFuzzer-tmp is increased by 96.2\% on average and the speed of SNPSFuzzer is increased by 140.45\% on average compared with AFLNET. We can also find that SNPSFuzzer is faster than SNPSFuzzer-tmp.
    Because SNPSFuzzer always selects the longer prefix messages, there are fewer messages left for the same message chain. Therefore, it can save the overhead of sending some messages and waiting for reply messages in each iteration of network protocol fuzzing to generate more test cases in the same time.
    
\begin{table}[htbp]
\centering
\caption{The Execution Speed of Each Fuzzer within 24 hours}
\label{tab:experiment_speed}
\resizebox{0.48\textwidth}{!}{%
\begin{tabular}{cccccc}
\hline
Program  & AFLNET &SNPSFuzzer-tmp  & SNPSFuzzer & $R_T$ & $R_S$  \\ \hline
Tinydtls & 2.5    &4.2  &5.3   &68.0\% $\uparrow$  &112.0\% $\uparrow$ \\ \hline
Kamailio & 4.5    &10.1 &12.1  &124.4\% $\uparrow$ &168.9\% $\uparrow$ \\ \hline
Average  & -        &-    &-     &96.2\% $\uparrow$ &140.45\% $\uparrow$ \\ \hline
\end{tabular}%
}
\end{table}

\subsection{Messages Savings Evaluation (RQ2)}
\begin{figure}[!t]
    \centering
    \subfigure[DTLS-Tinydtls]{
    \includegraphics[width=0.22\textwidth]{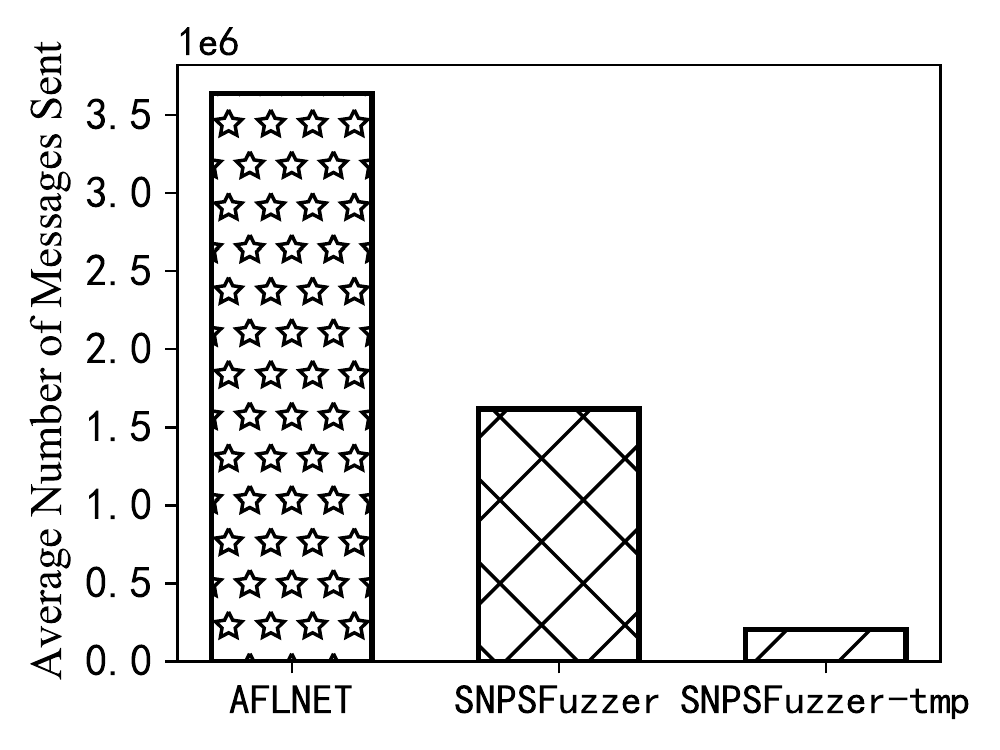}
    \label{messages_savings-a}
    }
    \subfigure[SIP-Kamailio]{
    \includegraphics[width=0.22\textwidth]{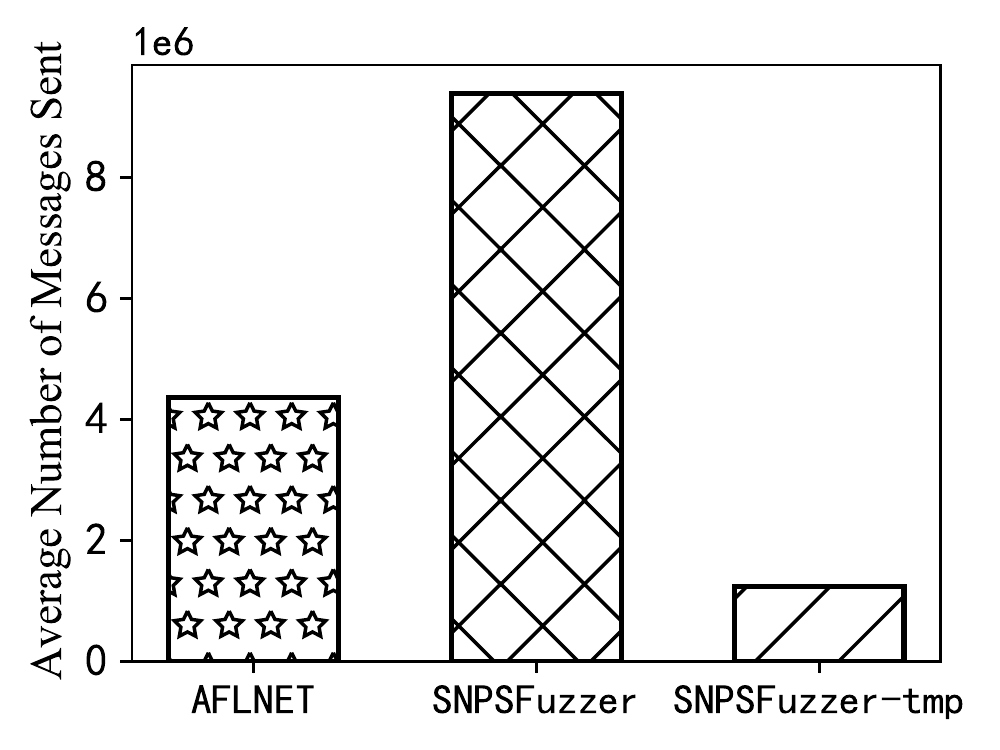}
    \label{messages_savings-b}
    }
    \caption{Average number of messages sent by AFLNET and average number of messages saved by SNPSFuzzer and SNPSFuzzer-tmp within 24 hours.}
    \label{fig:messages_savings}
    \end{figure}

    SNPSFuzzer uses snapshot technology to save the number of prefix messages sent. In the SCGF fuzzers represented by AFLNET, each test case contains multiple interrelated messages. Therefore, in order to better evaluate the performance of SNPSFuzzer, we counted the total number of messages sent by AFLNET and the number of messages saved by SNPSFuzzer and SNPSFuzzer-tmp in the same 24 hours. The results are shown in Figure \ref{fig:messages_savings}. 
    
    \textbf{Answer to RQ2.} We find that both SNPSFuzzer and SNPSFuzzer-tmp can save the number of messages sent, where SNPSFuzzer saves more messages. For program Tinydtls, SNPSFuzzer saves 44.4\% and SNPSFuzzer-tmp saves 5.6\% compared with the total number of messages sent by AFLNET. For program Kamailio, SNPSFuzzer even saves 215.4\% and SNPSFuzzer-tmp saves 28.4\% compared with the total number of messages sent by AFLNET. 
    
    There are two reasons for the above results: 1) The execution speed of SNPSFuzzer and SNPSFuzzer-tmp is faster than AFLNET. Therefore, SNPSFuzzer and SNPSFuzzer-tmp run more test cases in the same time. Compared with program Tintdtls, they saves more messages in program Kamailio; 2) SNPSFuzzer uses MCA algorithm to prioritize the selection deeper states. It needs to send more prefix messages to get the states. Therefore, SNPSFuzzer saves more messages than SNPSFuzzer-tmp.

\subsection{Code Coverage Evaluation (RQ3)}
    
\begin{figure}[!t]
    \centering
    \subfigure[DTLS-Tinydtls]{
    \includegraphics[width=0.22\textwidth]{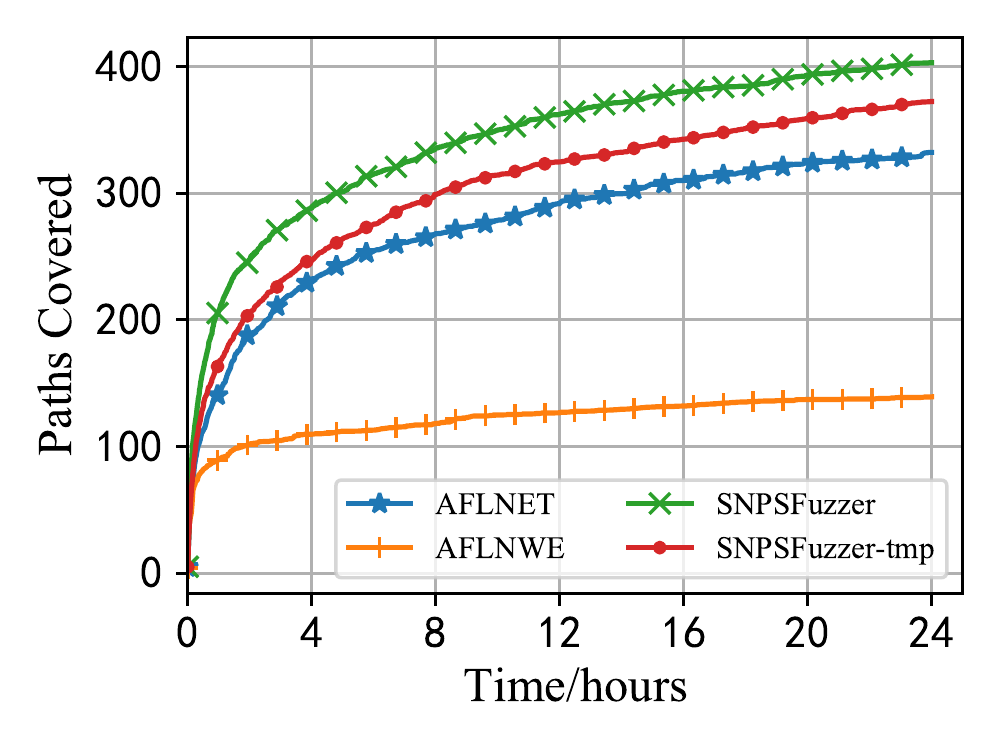}
    \label{path_coverage-dtls}
    }
    \subfigure[SIP-Kamailio]{
    \includegraphics[width=0.22\textwidth]{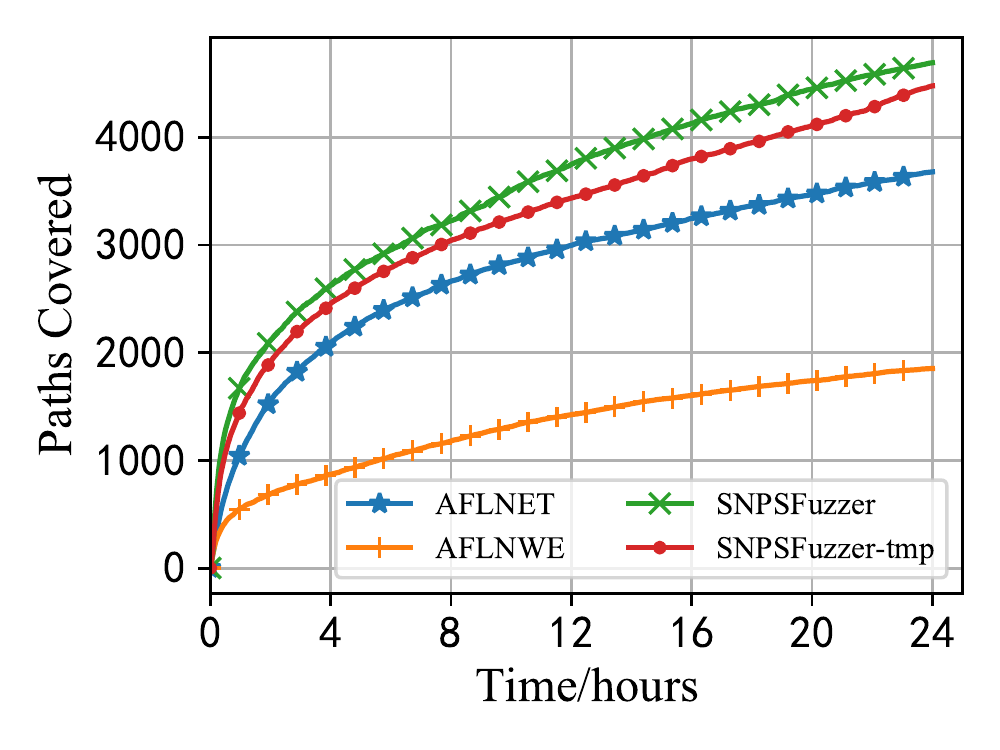}
    \label{path_coverage-sip}
    }
    \caption{Average number of paths covered by each fuzzer in 24 hours.}
    \label{fig:path_coverage}
    \end{figure}

    Figure \ref{fig:path_coverage} shows the detailed results of path covered changes during fuzzing each protocol and the overall improvements are summarized in Table \ref{tab:path_coverage}. 
    
    \textbf{Answer to RQ3.} We find that the efficiency of SNPSFuzzer in discovering new path coverage is significantly better than AFLNET and AFLNWE. At first, the number of path coverage found by each fuzzer increased rapidly. But with the passage of time, AFLNET gets into trouble faster and reach a state where it is difficult to grow rapidly. SNPSFuzzer can help alleviate this situation. Overall, in the same 24 hours, SNPSFuzzer can increase the path coverage by 24.45\% on average, of which at least 21.4\% can be increased. 
    
    We also find that SNPSFuzzer-tmp can improve the efficiency of path coverage discovery compared with AFLNET. In the same 24 hours, the found path coverage increased by 16.85\% on average. After using MCA algorithm, SNPSFuzzer can cover 7.6\% more new paths on average, which shows that MCA algorithm is helpful to cover deeper paths for network protocol fuzzing. 
    
    
    \begin{table}[htbp]
    \centering
    \caption{Average number of paths covered by each fuzzer in 24 hours}
    \label{tab:path_coverage}
    \resizebox{0.48\textwidth}{!}{%
    \begin{tabular}{ccccc}
    \hline
    Program  &AFLNET &AFLNWE & SNPSFuzzer-tmp & SNPSFuzzer   \\ \hline
    Tinydtls   &332  &139(58.1\%$\downarrow$)  &372(12.0\%$\uparrow$)   &403(21.4\%$\uparrow$)    \\ \hline
    Kamailio  &3680 &1854(49.6\%$\downarrow$)   & 4478(21.7\%$\uparrow$)  & 4693(27.5\%$\uparrow$)  \\ \hline
    Average &-  &53.85\% $\downarrow$    &16.85\% $\uparrow$ &24.45\%  $\uparrow$                    \\ \hline
    \end{tabular}%
    }
    \end{table}

\subsection{Vulnerability Discovery Evaluation (RQ4)}

    In order to show the performance of SNPSFuzzer in vulnerability discovery, we compared two metrics: unique crashes and unique vulnerabilities.
    
    \textbf{Answer to RQ4.} Table \ref{tab:unique_crashes} shows the number of unique crashes found by different fuzzers in the same 24 hours. We find that SNPSFuzzer finds more unique crashes than AFLNET with a total increase of 133.3\% and SNPSFuzzer-tmp finds more unique crashes than AFLNET with an average increase of 22.2\%. 

    SNPSFuzzer-tmp only uses snapshot technology to speed up network protocol fuzzing and thus vulnerabilities can be found faster. However, the results are limited. On the basis of snapshot technology, MCA algorithm can explore deeper protocol states without increasing the time overhead of sending prefix messages. For stateful network programs, messages parsing and abnormal state transition are key factors leading to vulnerabilities. Exploring the deep states helps to find the above two types of vulnerabilities. Therefore, MCA algorithm has a significant improvement in finding vulnerabilities.

    \begin{table}[htbp]
    \centering
    \caption{Average number of unique crashes found by different fuzzers in 24 hours}
    \label{tab:unique_crashes}
    \resizebox{0.48\textwidth}{!}{%
    \begin{tabular}{cccccc}
    \hline
    Program  & AFLNET &AFLNWE & SNPSFuzzer-tmp & SNPSFuzzer   \\ \hline
    Tinydtls   &34  & 24(29.4\% $\downarrow$)      &36(5.9\% $\uparrow$)  &38(11.8\% $\uparrow$)    \\ \hline
    Kamailio  & 2   & 0(100\% $\downarrow$)    &8(300\% $\uparrow$)    & 46(2200\% $\uparrow$)       \\ \hline
    Total  & 36 &  24(33.3\% $\downarrow$)   &44(22.2\% $\uparrow$)      &84(133.3\% $\uparrow$)            \\ \hline
    \end{tabular}%
    }
    \end{table}
    
    We further use Address Sanitizer\cite{addresssanitizer} to recompile the programs and reevaluate them with the discovered unique crash inputs. Finally, we get the number of unique vulnerabilities in program Tinydtls by analyzing the unique crashes. The results are shown in Table \ref{tab:bugs analysis}. SNPSFuzzer is superior to AFLNWE and AFLNET in the total number of unique vulnerabilities found. In the same 24 hours, compared with AFLNET, SNPSFuzzer can find a new unknown stack-buffer-overflow vulnerability. We have submitted the discovered vulnerability to the corresponding developers\cite{new_bug}.
    \begin{table}[htbp]
    \centering
    \caption{Statistics On Different Vulnerabilities in Tinydtls Exposed By Different Fuzzers in 24 hours}
    \label{tab:bugs analysis}
    \resizebox{0.48\textwidth}{!}{%
    \begin{tabular}{ccccc}
    \hline
    Program  & Vulnerability      &AFLNWE     & AFLNET & SNPSFuzzer \\ \hline
    Tinydtls & stack-buffer-overflow-1 &\checkmark &  \checkmark & \checkmark          \\
    & stack-buffer-overflow-2 &\checkmark &\checkmark &   \checkmark         \\
    & stack-buffer-overflow-3 &\XSolidBrush & \checkmark  &  \checkmark     \\
     & stack-buffer-overflow-4 &\XSolidBrush & \XSolidBrush  & \checkmark      \\ \hline
    Total    & 4             & 2/4          & 3/4    & 4/4        \\ \hline
    \end{tabular}%
    }
    \end{table}
    
    \textbf{Previous unknown vulnerability in program Tinydtls.} Listing \ref{new_bug} illustrates the unknown stack-buffer-overflow vulnerability exposed by SNPSFuzzer in Tinydtls. The location of the vulnerability is in Line 3776 of file $dtls.c$. When the program crashes, the value of $rlen$ is not 0, resulting in the increase of pointer $msg$ (Line 3912) until the program stack overflows. The value of $rlen$ is the return value of the function $is\_record$. The function $is\_record$ checks if the variable $msg$ points to a valid record of DTLS protocol and sets the sum of DTLS record layer and UDP layer length to variable $rlen$. If the value of $rlen$ is smaller than the value of $msglen$ (i.e. the length of received messages), $rlen$ can not be set 0 (Line 370-Line 371). SNPSFuzzer takes a snapshot after sending prefix messages, the mutated messages are added based on prefix messages. Therefore, SNPSFuzzer is more likely to generate long messages and quickly trigger the vulnerability. 
    \\
    
    \begin{lstlisting}[language=C, caption=Previously unknown vulnerability exposed by SNPSFuzzer in program Tinydtls., label = new_bug]
./dtls.c
356  static unsigned int is_record(uint8_t *msg,        size_t msglen) {
...
366    rlen = DTLS_RH_LENGTH + dtls_uint16_to_int(           DTLS_RECORD_HEADER(msg)->length);
...
370    if (rlen > msglen)	
371	      rlen = 0;
...
3756  int dtls_handle_message(dtls_context_t *ctx,     session_t *session, uint8_t *msg, int msglen)
... 
3776    while ((rlen = is_record(msg, msglen))) {
...
3912    			msg += rlen;
3913    			msglen -= rlen;
3914   			}  
\end{lstlisting}

\section{Related Work}

\subsection{Stateful Blackbox Fuzzing}
    The Stateful Blackbox Fuzzing (SBF) technology is widely used in network protocol fuzzing. Stateful blackbox fuzzing\cite{peach},\cite{boofuzz} usually uses the generation-based method to generate test cases without any program analysis. Specifically, in order to better explore the different state space of network protocol programs, SBF usually defines a set of template specifications in advance according to specific formats. For example, Peach provides an input template standard in XML format. SBF generates test cases that meet the protocol grammar according to the template file. The method brings good scalability to SBF and generally supports the fuzzing of a variety of network protocol programs. Whenever a new protocol needs to be fuzzed, just write the corresponding template file as required.

    However, the stateful blackbox fuzzing has two shortcomings. The detailed descriptions are as follows:
    
    One is that the final fuzzing results are closely related to the quality of the template file. Users need to write corresponding template files to describe the message formats and state transition relationship of the protocol after reading the RFC documents or other specifications. Therefore, in order to ensure the quality of fuzzing, users often need to spend a lot of time for writing template files. 
    
    The other one is that SBF can not use any program information to improve the quality of seeds. SBF is a blackbox fuzzing which can only judge whether the test case is valid by the external results of the program (e.g. crashes and timeout), and can not use other finer grained information. Therefore, the blackbox fuzzing does not save test cases even if these test cases generated in a fuzzing process are good for finding bugs. Therefore, blackbox fuzzing is often blind.
    
\subsection{Coverage-based Greybox Fuzzing}
    Coverage-based Greybox fuzzing (CGF) has attracted extensive attention recently. CGF obtains the coverage information of programs through lightweight instrument and uses genetic algorithm to improve the quality of test cases. CGF has achieved great success in fuzzing stateless programs such as command-line, file, library API programs. AFL\cite{afl} and LibFuzzer\cite{libFuzzer} are typical representatives.
    
    In recent years, most researchers have proposed various improved strategies and algorithms for CGF. MOPT\cite{mopt} proposes to use particle swarm optimization (PSO) algorithm to find the best probability distribution of CGF mutation operators. ENTROPIC\cite{boosting} develops a power schedule based on entropy for CGF to allocate more energy to the seeds that maximize information from the perspective of information theory. TortoiseFuzz\cite{TortoiseFuzz} proposes coverage accounting to address the limitation that CGF treats coverage equally.
    
    Moreover, many work on integrating the traditional static analysis and dynamic analysis technology into CGF is proposed. GREYONE\cite{greyone} introduces Dynamic Taint Analysis (DTA) technology into CGF and uses program data flow information to guide mutation. QSYM\cite{qsym} designs a fast concolic execution engine to support hybrid fuzzing. ParmeSan\cite{parmesan} presents sanitizer-guided fuzzing to reduce the Time-To-Exposure (TTE) of bugs.
    
    However, the above researches are mainly aimed at stateless programs, which can not effectively fuzz the stateful programs such as network programs.

\subsection{Stateful Coverage-based Greybox Fuzzing}
    In order to overcome the limitation of CGF in network protocol fuzzing, AFLNET\cite{aflnet} adds the concept of state to CGF. AFLNET is the first greybox fuzzer for protocol implementations, which realizes the network protocol programs fuzzing by sensing the message structures and state transition relationship of network protocol. Peach*\cite{9218603} is a greybox fuzzer for ICS protocol, which can identify the seeds that trigger the new code coverage and apply semantic sensitive generation strategy to optimize the generation of test cases. PAVFuzz\cite{PAVFuzz} can fuzz network protocols in the field of automatic driving such as RTPS (Real Time Publish Subscribe). It adopts a new relationship learning strategy, which can automatically learn the relationship between data elements in different protocol states. Using the learned relationship, PAVFuzz intelligently identifies the key elements of each state in the data model and dynamically assigns weights. Compared with the traditional random element selection strategy, dynamic weight avoids the waste of time and computing resources and thus significantly improves the efficiency of fuzzing.
    
    The above fuzzers have the function of network protocol state perception, and use the coverage information to improve the quality of test cases. However, they did not pay attention to the low efficiency caused by frequently sending the same prefix messages in the process of network protocol fuzzing. Therefore, this paper proposes to accelerate the network protocol fuzzing based on snapshot technology.
\section{Conclusion}
In this paper, we summarize the idea and process of existing SCGF technology in network protocol fuzzing, and point out that the speed of existing SCGF technology is generally slow. In order to speed up network protocol fuzzing, we propose SNPSFuzzer, a fast greybox fuzzer for stateful network protocol using snapshots. SNPSFuzzer dumps the context information of each state of the network protocol and restores it when the state needs to be fuzzed. Furthermore, on the basis of the snapshots technology, we design a message chain analysis algorithm to increase the number of prefix messages and explore more and deeper states in network protocol fuzzing. In the same 24 hours, SNPSFuzzer can generally increase the speed of network protocol fuzzing by 112.0\%-168.9\% and improve path coverage by 21.4\%-27.5\%. It has also found an unknown vulnerability in a real network program.
In future work, we will expand SNPSFuzzer to fuzz TCP based network protocols, such as Real Time Streaming Protocol (RTSP) and File Transfer Protocol (FTP). 



\bibliographystyle{IEEEtran}
\bibliography{reference}

\end{document}